\documentclass[prl,floats,twocolumn]{revtex4-1}
\usepackage{amsmath}
\usepackage{bbm}
\usepackage{graphicx}
\usepackage{color}

\begin{document}

\title{Marginal Fermi liquid versus excitonic instability in 3D Dirac semimetals}
\author{J. Gonz\'alez}
\address{Instituto de Estructura de la Materia,
        Consejo Superior de Investigaciones Cient\'{\i}ficas, Serrano 123,
        28006 Madrid, Spain}

\date{\today}

\begin{abstract}
We study the different phases in the Quantum Electrodynamics of 3D Dirac semimetals depending on the number $N$ of Dirac fermions, using renormalization group methods and the self-consistent resolution of the Schwinger-Dyson equation. We find that, for $N < 4$, a phase with dynamical generation of mass prevails at sufficiently strong coupling, sharing the same physics of the excitonic instability in 2D Dirac semimetals. For $N \geq 4$, we show that the phase diagram has instead a line of critical points characterized by the suppression of the quasiparticle weight at low energies, making the system to fall into the class of marginal Fermi liquids. Such a boundary marks the transition to a kind of strange metal which can be still defined in terms of electron quasiparticles, but with parameters that have large imaginary parts implying an increasing deviation from the conventional Fermi liquid picture.
\end{abstract}

\maketitle



{\em Introduction.---}
The discovery of graphene has opened new avenues of research in both theoretical and applied physics. The remarkable electronic properties of the material come to a great extent from the peculiar conical dispersion of the electron quasiparticles, which endows them with an additional pseudospin quantum number\cite{rmp}. This is an example of so-called Dirac semimetal, which provides an ideal playground to test many of the properties typical of relativistic fermion fields, like the Klein paradox\cite{kats} or the anomalous screening of charged particles\cite{nil,fog,shy,ter}.       

Another remarkable feature of relativistic field theories is the running of the coupling constants with the energy scale of the processes. In this regard, Dirac semimetals like graphene are described by a kind of Quantum Electrodynamics (QED) in two spatial dimensions. While the electron charge remains invariant in such a theory, the effective interaction strength is not constant however, as a result of the dependence of the Fermi velocity of electron quasiparticles on the energy scale. This behavior has been actually measured in suspended graphene samples at low doping levels\cite{exp2}, showing that the Fermi velocity grows as predicted\cite{np2,prbr} when looking close to the Dirac point (the vertex of the conical dispersion).

The effective interaction strength is thus progressively reduced in graphene at low energies, which may explain the absence of significant electronic correlations in the carbon layer. The recent discovery of materials with linear electronic dispersion in three dimensions\cite{wang1,wang2,liu,neupane,borisenko,yu} opens however the possibility of finding more exotic behaviors, stemming from the properties of QED in such a higher dimension. In the relativistic theory, the electron charge $e$ is screened at long distances by electron-hole pairs so that its value runs with the energy scale $\mu $, being related to the bare charge $e_\Lambda $ at the high-energy cutoff $\Lambda $ through the expression\cite{landau}
\begin{equation}
e^2_\Lambda = \frac{e^2 (\mu)}{1  -  \frac{1}{6\pi^2 c} e^2 (\mu) \log \frac{\Lambda}{\mu}  }
\label{pole}
\end{equation}
For constant $e_\Lambda $, this implies that the measurable charge $e (\mu )$ flows towards zero at low energies, leading to a weak-coupling regime in which we currently find the theory (with $e^2/4\pi c  \approx  1/137$).
 
Assuming conversely that $e$ has some finite value at energy $\mu $, the above equation shows that the bare coupling $e_\Lambda $ should blow up at a certain value of the large cutoff $\Lambda $. This is the well-known Landau pole\cite{landau}, that for some time cast many doubts about the quantum field theory approach to the description of elementary particles, given the impossibility to attach any physical meaning to such a high-energy singularity. In the condensed matter context, however, the high-energy cutoff $\Lambda $ is a magnitude that can be related to the short-distance scale of the microscopic lattice, making sense to ask about the influence of the Landau pole or, more generically, the effect of the scaling of the electron charge in the QED of 3D Dirac semimetals. This is a relevant question to address the physics of materials naturally placed in a regime of strong $e$-$e$ interaction, for which the effective strength is given in general by the ratio between $e^2$ and the Fermi velocity $v_F$ of the electron quasiparticles. 

In the present paper, we investigate the different phases in the QED of 3D Dirac semimetals, in which the speed of light $c$ is replaced by a much smaller Fermi velocity $v_F$. Taking formally the limit of a large number $N$ of fermion flavors, we will see that such a theory has a critical point in the effective interaction strength $g \equiv Ne^2/2\pi^2 v_F $ at $g_c = 3$. This critical value will be obtained for the renormalized coupling arising in a rigorous scale-invariant calculation of the electron scaling dimension, showing the vanishing of the electron quasiparticle weight at the critical point. A similar result will be also found by the self-consistent resolution of the Schwinger-Dyson equation for the electron propagator, allowing to obtain a definite picture of marginal Fermi liquid behavior from the renormalization of the quasiparticle weight.                

\begin{figure}[h]
\begin{center}
\includegraphics[width=3.9cm]{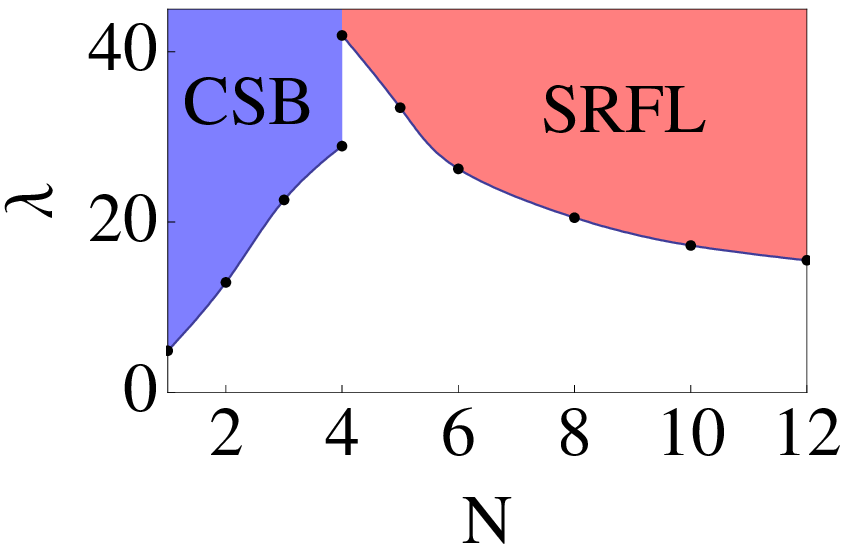}
\hspace{0.2cm}
\includegraphics[width=4.3cm]{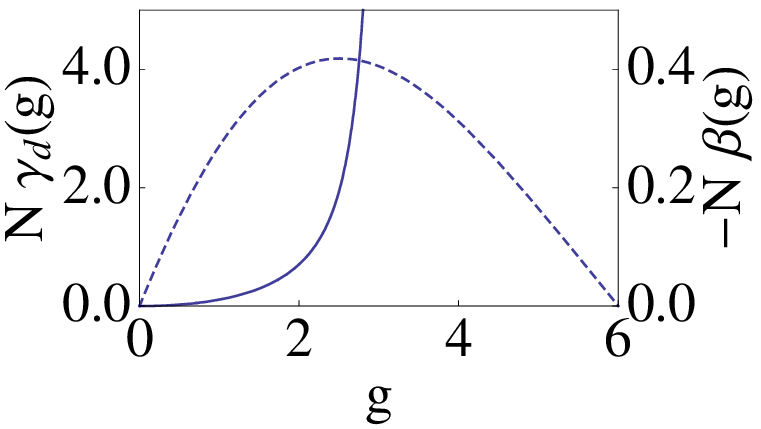}\\
 \hspace{0.36cm}  (a) \hspace{3.6cm} (b)
\end{center}
\caption{(a) Phase boundaries given in terms of the bare interaction strength $\lambda $ (defined in Eq. (\ref{def})), marking the transition to a strongly renormalized Fermi liquid (SRFL) at large $N$ and to a phase with chiral symmetry breaking (CSB) for small $N$. (b) Plot of the anomalous scaling dimension $\gamma_d (g)$ (full line) and the rate of variation $\beta (g)$ of the Fermi velocity with respect to energy (dashed line), multiplied both by $N$.}
\label{one}
\end{figure}

The resolution of the Schwinger-Dyson equation will make also possible to identify the phases of the system when $N$ is not large, leading to two different boundaries in the complete phase diagram shown in Fig. \ref{one}(a). Thus, we will see that for $N < 4$ there is a transition to a phase with dynamical generation of mass at sufficiently strong coupling. The case with $N = 4$ is special in that the chiral symmetry breaking turns out to be assisted by the own vanishing of the quasiparticle weight. For $N > 4$, we will find that the critical line corresponds to the mentioned marginal Fermi liquid behavior, marking the transition to a kind of ^^ ^^ strange metal" which can be still defined in terms of electron quasiparticles, but with parameters that get large imaginary parts implying an increasing deviation from the Fermi liquid behavior.

{\em Scaling properties of 3D Dirac semimetals.---}
We describe the QED of Dirac fermions with Fermi velocity $v_F \ll c$ starting from the hamiltonian for a collection of $N$ four-component Dirac spinors $\{ \psi_i \}$ in generic spatial dimension $D$
\begin{equation}
H = i v_F \int d^D r \; \psi^{\dagger}_i({\bf r}) 
 \gamma_0 \mbox{\boldmath $\gamma $}  \cdot  \mbox{\boldmath $\nabla $} \psi_i ({\bf r})   
     + e_0 \int d^D r  \psi^{\dagger}_i({\bf r})  \psi_i ({\bf r})  \phi ({\bf r})
\label{ham}
\end{equation}
where $\phi ({\bf r})$ stands for the scalar potential and $\{ \gamma_\alpha \}$ is a set of Dirac matrices satisfying $\{\gamma_\alpha , \gamma_\beta \} = 2\eta_{\alpha \beta}$. Here $\eta $ represents the Minkowski metric $\eta = {\rm diag}(-1,1,\ldots 1)$, so that the kinetic term in the hamiltonian has eigenvalues $\pm v_F |{\bf k}|$ in momentum space. The physical dimension corresponds to $D = 3$, but we will start shifting it formally to $D = 3 - \epsilon $ in order to regularize the divergences that the theory has at large momenta, following a procedure aimed to preserve the gauge invariance in the computation of observable quantities\cite{juricic,jhep}.     

In the non-relativistic regime $v_F \ll c$, $\phi $ mediates the Coulomb interaction between electrons and it has a free propagator in momentum space $D_0({\bf q},\omega ) = e_0^2 /{\bf q}^2$. This is corrected by the electron-hole polarization $\Pi ({\bf q},\omega )$, which is a divergent quantity at $D = 3$\cite{iz}. Computing to leading order in a $1/N$ expansion, we get the expression for the $\phi $ propagator
\begin{equation}
D({\bf q},\omega ) = \frac{e_0^2}{{\bf q}^2 + N B(\epsilon ) \frac{e_0^2}{2\pi^2 v_F} \frac{{\bf q}^2}{(v_F^2 {\bf q}^2-\omega^2)^{\epsilon/2}}}
\label{iprop}
\end{equation}
with $B(\epsilon ) = (4\pi )^{\epsilon/2} \Gamma(\epsilon/2) \Gamma(2-\epsilon/2)^2 / \Gamma(4-\epsilon)$. The divergence as $\epsilon \rightarrow 0$ can be reabsorbed into a simple renormalization of the bare electron charge $e_0$, passing to the physical dimensionless coupling $e$ with the help of an auxiliary energy scale $\mu $ through the redefinition $\mu^\epsilon /e_0^2 = 1/e^2 - N/6\pi^2 v_F \epsilon $. We have then
\begin{equation}
e_0^2 = \frac{\mu^\epsilon e^2}{1 - \frac{N}{6\pi^2 v_F} e^2 \frac{1}{\epsilon }}
\end{equation}
which is the counterpart of Eq. (\ref{pole}) in the dimensional regularization approach, where the $\log (\Lambda)$ dependence is replaced by the $1/\epsilon $ pole\cite{f1}.

We end up in this way with an expression of the $\phi $ propagator which is finite in the limit $\epsilon \rightarrow 0$,
\begin{equation}
D({\bf q},\omega ) = \frac{\mu^\epsilon e^2}{{\bf q}^2 \left(1 - \frac{Ne^2}{6\pi^2 v_F} \frac{1}{\epsilon }  + N B(\epsilon ) \frac{e^2}{2\pi^2 v_F} \frac{\mu^\epsilon}{(v_F^2 {\bf q}^2-\omega^2)^{\epsilon/2}} \right)}
\label{prop}
\end{equation}   
For the computation of different observable quantities, one has still to keep however a nonzero $\epsilon $. A crucial property of the theory is the so-called renormalizability, by which physical quantities turn out to be finite in the limit $\epsilon \rightarrow 0$ and, moreover, with no dependence on the auxiliary scale $\mu $. We have checked that these conditions are met to leading order in the $1/N$ expansion (see Appendix), assuring that the different scaling dimensions only depend on the renormalized coupling $e$ (and the renormalized Fermi velocity).


We study then the effect of quantum corrections on the electron quasiparticle properties. For 
that purpose, one can compute the electron self-energy $\Sigma ({\bf k},\omega_k )$, which is given to leading order of the $1/N$ expansion by
\begin{equation}
i \Sigma ({\bf k},\omega_k ) = -\int \frac{d^D p}{(2\pi )^D} \frac{d\omega_p }{2\pi } G_0({\bf k}-{\bf p}, \omega_k - \omega_p)  D({\bf p},\omega_p )
\end{equation}
where $G_0({\bf p}, \omega_p)$ stands for the free Dirac propagator. The self-energy develops its own divergences in the limit $\epsilon \rightarrow 0$, which can be completely absorbed into a redefinition of quasiparticle parameters by renormalization factors $Z_\psi$ and $Z_v$ in the expression of the propagator
\begin{equation}
G ({\bf k},\omega_k )^{-1}   =  Z_\psi (\omega_k  -  Z_v v_F  \gamma_0 \mbox{\boldmath $\gamma $}  \cdot  {\bf k}  )  - Z_\psi \Sigma ({\bf k},\omega_k )            
\end{equation}

The renormalization factors are functions of the effective coupling $g = N e^2/2\pi^2 v_F$, having the pole structure $Z_\psi = 1 + (1/N)\sum_{n=1}^{\infty } c_n (g)/\epsilon^n , Z_v = 1 + (1/N)\sum_{n=1}^{\infty } b_n (g)/\epsilon^n$. In the present theory, the electron propagator $G$ can be made free of poles in the $\epsilon $ variable with an appropriate choice of coefficients $c_n (g)$ and $b_n (g)$ which do not depend on the auxiliary scale $\mu $ (see Appendix). This is a crucial property, since the electronic correlators get anomalous scaling dimensions\cite{amit} that are given by multiples of  
\begin{equation}
\gamma_d = \frac{\mu }{Z_\psi } \frac{\partial Z_\psi }{ \partial \mu }
\end{equation}
Under the rescaling ${\bf k} \rightarrow s{\bf k}$, $\omega \rightarrow s\omega $, the electron propagator becomes for instance 
\begin{equation}
G(s{\bf k}, s\omega ) \approx  s^{-1 + \gamma_d }  G({\bf k}, \omega )
\label{anom}
\end{equation}
In the present case, the only dependence of $Z_\psi $ on $\mu $ comes from the dependence implicit in the coupling $g$, leading to  $\gamma_ d (g) = - g \: c_1'(g)/N $ \cite{amit}. This allows to obtain such an observable quantity exclusively in terms of the physical coupling $g$.   

We have computed the coefficient $c_1(g)$ up to very high orders in the coupling $g$, finding that these approach a precise geometric sequence. This means that the power series in $g$ has a finite radius of convergence, which we have determined to be at $g_c = 3$ (see Appendix). The main consequence of this behavior of $c_1(g)$ is the divergence of the anomalous exponent $\gamma_d$ at such a critical coupling, as displayed in Fig. \ref{one}(b). According to (\ref{anom}), this has to be interpreted as the suppression of the electron quasiparticle weight in the low-energy limit.

One can check that the renormalized Fermi velocity has instead a regular behavior at the critical point. This can be seen from inspection of the residues of the poles in $Z_v $, that remain finite at $g_c$. The condition of independence of the bare Fermi velocity on the auxiliary scale, $\mu \partial (Z_v v_F) / \partial \mu = 0$, leads to a scaling equation for the renormalized Fermi velocity
\begin{equation}
\frac{\mu }{v_F }  \frac{\partial v_F}{\partial \mu }  = \beta (g)
\end{equation}
with $\beta (g) = g b_1'(g)/N$ (see Appendix). This is a negative bounded function up to $g_c$, as seen in Fig. \ref{one}(b), giving rise therefore to a limited growth of the Fermi velocity in the low-energy limit $\mu \rightarrow 0$ \cite{f2}. We conclude then that the singularity found at the critical coupling does not produce a qualitative change in the electronic dispersion, but rather translates into a strong attenuation of the own electron quasiparticles.

While the critical point is found at the coupling $g_c$, we have to bear in mind that such a critical value refers to a renormalized coupling that has an implicit dependence on the energy scale of the type shown in Eq. (1). The above scale-invariant calculation of $g_c$ does not give however any indication about the particular energy at which the critical coupling is obtained. This ambiguity can be overcome by dealing instead with a computational procedure that keeps memory of the high-energy cutoff, allowing for instance to refer the critical point to measurable parameters of the bare theory defined at short-distance scales, as we illustrate in the next section.


{\em Beyond the large-$N$ approximation.---}
In order to access the phases of the electron system at low values of $N$, we require an approach with a more comprehensive sum of many-body corrections, beyond those considered in the large-$N$ approximation. With this aim, we adopt next an alternative approach consisting in the self-consistent resolution of the Schwinger-Dyson equation for the electron propagator, that amounts to include all kind of diagrammatic contributions except those containing vertex corrections. To make the comparison with the results in the previous section, we characterize the quasiparticle properties in terms of the functions $z_\psi({\bf k}, \omega), z_v({\bf k}, \omega)$ and $z_m({\bf k}, \omega)$, writing the electron propagator in the form
\begin{equation}
G({\bf k}, \omega)  =  \left(  z_\psi({\bf k}, \omega) \omega  -  z_v({\bf k}, \omega) v_F \gamma_0 \mbox{\boldmath $\gamma $}  \cdot  {\bf k}  -  z_m({\bf k}, \omega) \gamma_0 \right)^{-1}
\end{equation}
The function $z_m({\bf k}, \omega)$ is now introduced to study the possible dynamical generation of a mass term (and the consequent opening of a gap at the Dirac point) assuming that the original theory does not have such a bare coupling in the hamiltonian. 

The resolution proceeds by computing the propagator $D({\bf q},\omega )$ as in Eq. (\ref{iprop}), but taking now the polarization 
\begin{equation}
i\Pi ({\bf q},\omega_q ) =  \int \frac{d^3 p}{(2\pi )^3} \frac{d\omega_p }{2\pi } {\rm Tr} \left[ G({\bf q}-{\bf p}, \omega_q - \omega_p)  G({\bf p},\omega_p )  \right]
\label{pi}
\end{equation}
The functions $z_\psi({\bf k}, \omega), z_v({\bf k}, \omega)$ and $z_m({\bf k}, \omega)$ must be then adjusted to attain self-consistency in the evaluation of the propagator $G({\bf k},\omega_k )$ corrected with the self-energy
\begin{equation}
i \Sigma ({\bf k},\omega_k ) = -\int \frac{d^3 p}{(2\pi )^3} \frac{d\omega_p }{2\pi } G({\bf k}-{\bf p}, \omega_k - \omega_p)  D({\bf p},\omega_p )
\label{si}
\end{equation}
In practice, these equations must be solved rotating all the frequencies in the complex plane, $\omega = i\overline{\omega} $, passing then to a Euclidean space in the variables $({\bf k}, \overline{\omega} )$.

The main difference with respect to the previous regularization is that now the computation of the integrals in (\ref{pi}) and (\ref{si}) requires the introduction of a high-momentum cutoff $\Lambda_k $. Accordingly, we may introduce a renormalized coupling $e (\mu )$ related to the charge $e_0$ at $\Lambda_k $ as in Eq. (\ref{pole}). We have to bear in mind however that the nominal parameters $e_0$ and $v_F$ used in the self-consistent resolution are not directly observable, as they may differ appreciably from the respective final quantities measurable at short-distance scales. In this respect, it is more sensible to define the bare charge $e_B$ in terms of the interaction propagator from the relation $e_B^2 = \Lambda_k^2 \: D(\Lambda_k,0)$. Similarly, we can define the bare Fermi velocity from the electron propagator as $v_B = z_v(\Lambda_k, 0) \: v_F$. The bare interaction strength can be then estimated from the coupling 
\begin{equation}
\lambda = N e_B^2/2\pi^2 v_B
\label{def}
\end{equation} 
In this approach, the electron charge $e_B$ must be set to its standard value, while $v_B$ can be taken as a variable (depending on the particular material) with which we can move $\lambda $ from weak to strong coupling.

\begin{figure}[h]
\begin{center}
\includegraphics[width=4.3cm]{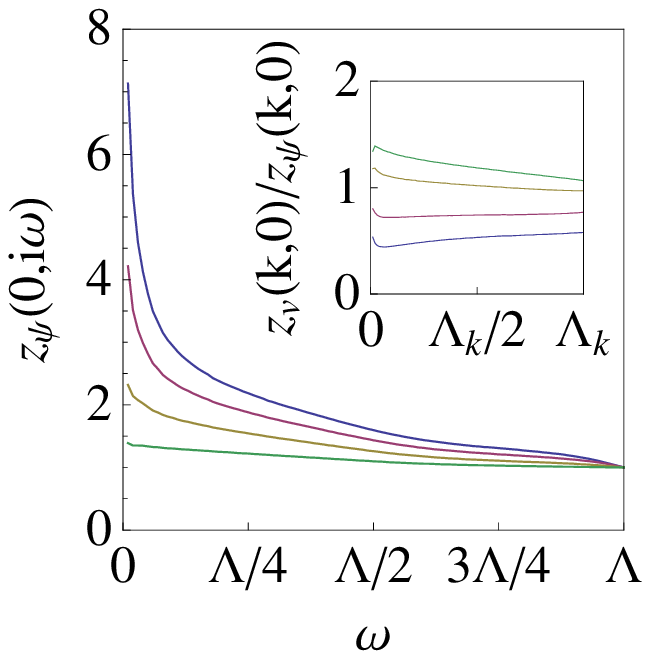}
\hspace{0.2cm}
\includegraphics[width=3.9cm]{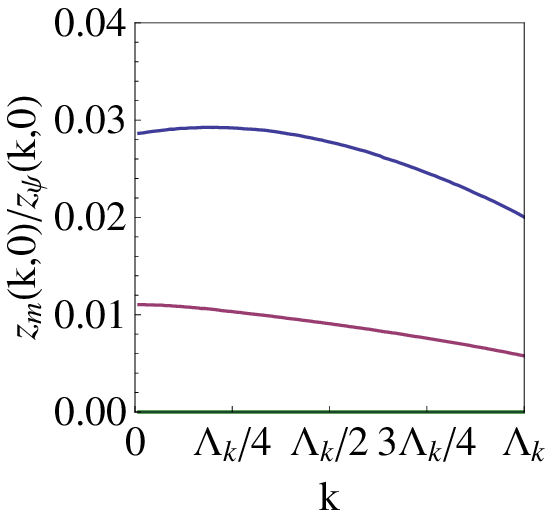}\\
 \hspace{0.36cm}  (a) \hspace{3.6cm} (b)
\end{center}
\caption{Plot of the factors $z_\psi (0,i\omega)$ and $z_m ({\bf k},0)/z_\psi ({\bf k},0)$ (expressed in eV) for $N = 4$ and values of the bare interaction strength $\lambda = 42.4, 31.6, 19.4$ and $8.1$ (from top to bottom, with the two lowest curves in (b) collapsed down to the horizontal axis). The inset shows the plot of $z_v ({\bf k},0)/z_\psi ({\bf k},0)$ for the same sequence of couplings, from bottom to top.}
\label{two}
\end{figure}

Thus, for all values of $N$, we find first a phase connected to weak coupling with regular Fermi liquid behavior, corresponding to the regime where a purely real solution exists for $z_\psi({\bf k}, i\overline{\omega})$ and $z_v({\bf k}, i\overline{\omega})$, while $z_m({\bf k}, i\overline{\omega})$ turns out to be self-consistently set to zero. For all $N \geq 4$, we find moreover a critical point $\lambda_c$, characterized by the divergence of $z_\psi({\bf k}, i\overline{\omega})$ in the limit of vanishing frequency together with a soft renormalization of the Fermi velocity $z_v({\bf k}, i\overline{\omega})/z_\psi({\bf k}, i\overline{\omega})$, as shown in Fig. \ref{two}(a). The critical behavior is then governed by the vanishing of the quasiparticle weight at low energies, in clear correspondence with the features found in the dimensional regularization of the theory. As shown in Fig. \ref{one}(a), the critical value $\lambda_c$ for the bare coupling approaches an asymptotic limit at large $N$, which is the counterpart of $g_c$ obtained for the renormalized coupling in the previous section.

The case with $N = 4$ is special however in that it also leads to a nonvanishing $z_m({\bf k}, i\overline{\omega})$ assisted by the own divergence of $z_\psi({\bf k}, i\overline{\omega})$, as evidenced in Fig. \ref{two}. The interpretation we can make is that, for this particular value of $N$, the dynamical generation of mass and the suppression of the quasiparticle weight reinforce each other, due to the consequent reduction in the screening of the Coulomb interaction. For $N < 4$, we find that the development of a nonvanishing $z_m({\bf k}, i\overline{\omega})$ clearly prevails, leading to the phase with chiral symmetry breaking mapped in Fig. \ref{one}(a).

\begin{figure}[h]
\begin{center}
\includegraphics[width=4.0cm]{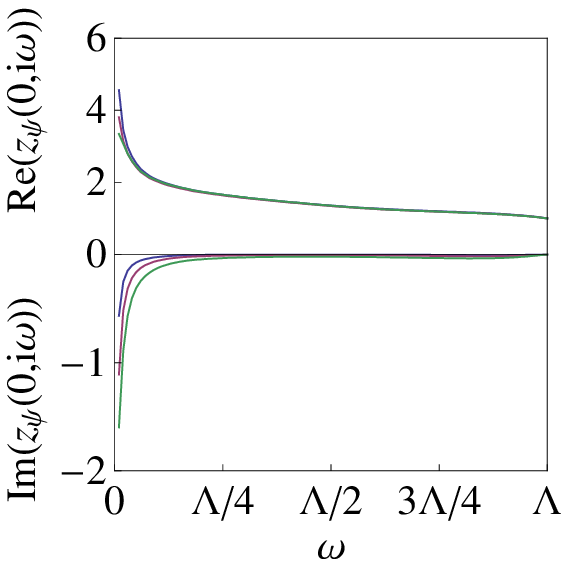}
\hspace{0.2cm}
\includegraphics[width=4.2cm]{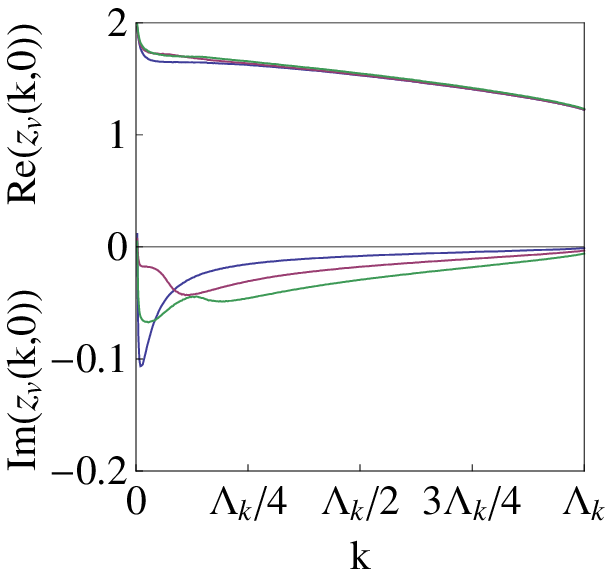}\\
 \hspace{0.36cm}  (a) \hspace{3.6cm} (b)
\end{center}
\caption{Plot of the real and imaginary parts of the renormalization factors $z_\psi (0,i\omega)$ and $z_v ({\bf k},0)$ for $N = 8$ and values of the bare interaction strength $\lambda = 21.08, 21.13$ and $21.77$ (curves from top to bottom in both sides of (a), from lower to higher absolute value in both sides of (b)).}
\label{three}
\end{figure}

Finally, a nice feature of the present approach is that it also allows to investigate the properties of the theory above the critical point for $N \geq 4$. The self-consistent resolution can be carried out when $\lambda > \lambda_c $ in the same fashion as before, with the result that $z_\psi({\bf k}, i\overline{\omega})$ and $z_v({\bf k}, i\overline{\omega})$ become now complex functions. We can ascribe this new behavior to the onset of a different phase of the electron system, in which electron quasiparticles still exist but with a decay rate dictated by the imaginary contributions in the self-energy. These may get very large at low energies, as shown in Fig. \ref{three}, meaning that we are dealing in this regime with a kind of ^^ ^^ strange" metal with very unstable quasiparticles whose decay rate does not vanish even at the Dirac point.

{\em Conclusion.---}
We have seen that, unlike 2D Dirac semimetals that have a dominant instability towards exciton condensation at strong coupling\cite{khves,gus,vafek,khves2,her,jur,drut1,hands2,gama,fer,ggg,prb}, their 3D analogues have a richer phase diagram with two different critical lines. The location of these phase boundaries can be understood from the interplay between the tendency to dynamical generation of mass, which is similar to the excitonic instability in 2D Dirac semimetals (and similar also to the chiral symmetry breaking of the fully covariant QED in 4D space-time\cite{mas,fom,fuk,mir,gus2,kon,atk,min}), and the suppression of the electron quasiparticle weight at low energies, that appears as the natural instability in 3D Dirac semimetals at large $N$. 

This latter behavior provides a genuine example within the class of so-called marginal Fermi liquids, introduced some years ago in the effort to understand the properties of the normal state of copper-oxide superconductors\cite{varma,bares,nayak,hou,cast}. The 3D Dirac semimetals offer now the possibility to access such a regime in materials with sufficiently small Fermi velocity. Taking the values represented in Fig. \ref{one}(a) (and having in mind the definition $\lambda = N e_B^2/2\pi^2 v_B$), it is easy to see that the critical point should be reached for instance with $N = 8$ for a bare Fermi velocity $v_B$ of about one half the typical value in a graphene sheet.


We have also investigated the phase arising when the effective coupling is larger than the critical coupling $\lambda_c$ for marginal Fermi liquid behavior. Our self-consistent resolution of the Schwinger-Dyson equations has led us to predict then a ^^ ^^ strange metal" phase with very unstable quasiparticles and deviating increasingly from the Fermi liquid picture, providing definite signatures susceptible of being confirmed by the experimental observation of 3D Dirac semimetals with sufficiently small Fermi velocity.

The financial support from MICINN (Spain) through grant FIS2011-23713 is gratefully acknowledged.

\onecolumngrid

\vspace{3cm}

\begin{center}
{\bf APPENDIX}
\end{center}

\begin{center}
{\bf Evaluation of the critical coupling of 3D Dirac semimetals in $1/N$ expansion}
\end{center}

The electron self-energy computed to leading order in the $1/N$ expansion is given by the expression
\begin{equation}
i\Sigma ({\bf k},\omega_k ) = -\int \frac{d^D p}{(2\pi )^D} \frac{d\omega_p }{2\pi } \:
  \frac{\omega_k - \omega_p + v_F \gamma_0 \mbox{\boldmath $\gamma $}  \cdot  ({\bf k} - {\bf p})}
           {(\omega_k - \omega_p)^2 - v_F^2 ({\bf k} - {\bf p})^2 + i\delta}  \:
 \frac{\mu^\epsilon e^2}{{\bf p}^2 \left(1 - \frac{Ne^2}{6\pi^2 v_F} \frac{1}{\epsilon }  + N B(\epsilon ) \frac{e^2}{2\pi^2 v_F}
       \frac{\mu^\epsilon}{(v_F^2 {\bf p}^2-\omega_p^2)^{\epsilon/2}} \right)}     
\end{equation}
with $B(\epsilon ) = (4\pi )^{\epsilon/2} \Gamma(\epsilon/2) \Gamma(2-\epsilon/2)^2 / \Gamma(4-\epsilon)$. In order to obtain the real part of $\Sigma ({\bf k},\omega_k )$, we can perform a rotation to imaginary frequencies $\omega = i\overline{\omega }$. The self-energy becomes then, as a function of the effective coupling $g = N e^2/2\pi^2 v_F$,
\begin{equation}
\Sigma ({\bf k},i\overline{\omega}_k ) = \frac{2\pi^2}{N} v_F  \int \frac{d^D p}{(2\pi )^D} \frac{d\overline{\omega}_p }{2\pi }  \:
  \frac{i\overline{\omega}_k - i\overline{\omega}_p + v_F \gamma_0 \mbox{\boldmath $\gamma $}  \cdot  ({\bf k} - {\bf p})}
           { v_F^2 ({\bf k} - {\bf p})^2 + (\overline{\omega}_k - \overline{\omega}_p)^2}  \:
    \frac{g \mu^\epsilon }{{\bf p}^2 \left(1 - g \frac{1}{3\epsilon }  + g B(\epsilon )
       \frac{\mu^\epsilon}{(v_F^2{\bf p}^2+\overline{\omega}_p^2)^{\epsilon/2}} \right)}     
\end{equation}

In the analytic continuation to dimension $D = 3-\epsilon $, the divergences that the self-energy would develop as logarithms of the high-energy cutoff $\Lambda $ appear as powers of $1/\epsilon $. One has to check that these poles can be absorbed into appropriate renormalization factors $Z_\psi$ and $Z_v$, rendering convergent at $D = 3$ the electron propagator $G ({\bf k},\omega_k )$ given by
\begin{equation}
G ({\bf k},\omega_k )^{-1}   =  Z_\psi (\omega_k  -  Z_v v_F  \gamma_0 \mbox{\boldmath $\gamma $}  \cdot  {\bf k}  )  - Z_\psi \Sigma ({\bf k},\omega_k )            
\end{equation}
The renormalization factors must have the pole structure
\begin{eqnarray}
Z_\psi  & = &  1 + \frac{1}{N}\sum_{n=1}^{\infty } \frac{c_n (g)}{\epsilon^n}      \\
Z_v  & = &  1 + \frac{1}{N}\sum_{n=1}^{\infty } \frac{b_n (g)}{\epsilon^n}
\end{eqnarray}
The residues $c_n (g)$ and $b_n (g)$ can be obtained in the form of power series in the $g$ coupling starting from the expansion
\begin{equation}
\Sigma ({\bf k},i\overline{\omega}_k ) = \frac{2\pi^2}{N} v_F \int \frac{d^D p}{(2\pi )^D} \frac{d\overline{\omega}_p }{2\pi }   \:
  \frac{i\overline{\omega}_k - i\overline{\omega}_p + v_F \gamma_0 \mbox{\boldmath $\gamma $}  \cdot  ({\bf k} - {\bf p})}
           { v_F^2 ({\bf k} - {\bf p})^2 + (\overline{\omega}_k - \overline{\omega}_p)^2}   \:
    \frac{\mu^\epsilon }{{\bf p}^2} \sum_{n=0}^{\infty} (-1)^n g^{n+1} \left(- \frac{1}{3\epsilon } + B(\epsilon )
       \frac{\mu^\epsilon}{(v_F^2{\bf p}^2+\overline{\omega}_p^2)^{\epsilon/2}} \right)^n     
\end{equation}

To compute for instance the renormalization of the quasiparticle weight, we can set ${\bf k} = 0$ and use systematically the result
\begin{eqnarray}
\lefteqn{  \int \frac{d^D p}{(2\pi )^D} \frac{d\overline{\omega}_p }{2\pi }   \: 
    \frac{i\overline{\omega}_k - i\overline{\omega}_p }
           { {\bf p}^2 + (\overline{\omega}_k - \overline{\omega}_p)^2}  \:
       \frac{1}{{\bf p}^2}   \frac{1}{({\bf p}^2+\overline{\omega}_p^2)^{m\epsilon/2}}   = }   \nonumber   \\
  &   &  i\overline{\omega}_k \frac{1}{(4\pi)^{2-\epsilon/2}} \frac{m\epsilon}{1-\epsilon}
   \frac{\Gamma(\frac{m+1}{2}\epsilon) \Gamma(1-\frac{m+1}{2}\epsilon) \Gamma(1-\frac{\epsilon}{2})}
     {\Gamma(1+\frac{m}{2}\epsilon) \Gamma(2-\frac{m+2}{2}\epsilon) }   \frac{1}{|\overline{\omega}_k|^{(m+1)\epsilon}}
\label{reslt}  
\end{eqnarray}
In this way we obtain the analytic expression of the first orders of the residues $c_n (g)$
\begin{eqnarray}
c_1 (g ) & = &  -  \frac{1}{24} g^2
    - \frac{1}{192}  g^3  -  \frac{5}{5184}  g^4 
    - \left( \frac{1}{6480}+\frac{\zeta (3)}{6480}  \right)  g^5 
   - \left( \frac{7}{279936}+\frac{\pi ^4}{2799360}+\frac{\zeta (3)}{34992} \right)
                 g^6                                           \nonumber     \\  
   &  &  - \left( \frac{1}{244944}+\frac{\pi ^4}{14696640}+\frac{5 \zeta (3)}{979776}
           +\frac{\zeta (5)}{108864}  \right)  g^7    +  \ldots    \label{cs0}    \\
c_2 (g ) & = &  - \frac{1}{108}  g^3  -  \frac{1}{648} g^4 -  \frac{1}{3888}  g^5  
  -  \left(  \frac{1}{23328}+\frac{\zeta (3)}{23328}  \right)  g^6  
  - \left(  \frac{1}{139968}+\frac{\pi ^4}{9797760}+\frac{\zeta (3)}{122472} \right) g^7  
                                    +  \ldots                                        \\
c_3 (g ) & = &  -  \frac{1}{432}  g^4  -  \frac{1}{2430} g^5 - \frac{5}{69984}  g^6  
   - \left( \frac{1}{81648}+\frac{\zeta (3)}{81648} \right) g^7 +  \ldots      \\
c_4 (g ) & = &  - \frac{1}{1620} g^5  -  \frac{1}{8748}   g^6  
                    - \frac{5}{244944} g^7    +  \ldots                      \\
c_5 (g ) & = &  -  \frac{1}{5832}  g^6  -  \frac{1}{30618}   g^7  +   \ldots    \\
c_6 (g ) & = &  -  \frac{1}{20412}  g^7  +   \ldots
\label{cs}
\end{eqnarray}
A most important feature regarding these expressions is that they do not contain any logarithmic dependence (in fact any dependence) on $\overline{\omega}_k$, which is crucial to guarantee the interpretation of $Z_\psi$ as the renormalization of a local operator in the original theory.

Moreover, another remarkable property of the residues $c_n (g)$ is that they lead to an anomalous scaling dimension $\gamma_d$ which is finite in the limit $\epsilon \rightarrow 0$. This can be shown by first realizing that the only dependence of $Z_\psi$ on the auxiliary scale $\mu $ comes through the renormalized coupling $g$. We can write therefore
\begin{equation}
\gamma_d = \frac{\mu }{Z_\psi } \frac{\partial g }{ \partial \mu } \frac{\partial Z_\psi }{ \partial g } 
\label{gammad} 
\end{equation}
The derivative of $g$ with respect to $\mu $ can be obtained by exploiting the independence of the parameters of the bare theory with respect to that energy scale. Differentiating the expression
\begin{equation}
\frac{1}{e_0^2} = \mu^{-\epsilon } \left( \frac{1}{e^2} - \frac{N}{6\pi^2 v_F} \frac{1}{\epsilon } \right)
\end{equation}
and passing to the variable $g = N e^2/2\pi^2 v_F$, we find
\begin{equation}
\mu \frac{\partial g }{ \partial \mu } = -\epsilon g + \frac{1}{3} g^2
\label{scalg}
\end{equation}
Using this result in (\ref{gammad}), we may write
\begin{equation}
\gamma_d = \frac{1}{Z_\psi } \frac{1}{N} \left( - g \sum_{n=0}^{\infty } \frac{c_{n+1}' (g)}{\epsilon^n} + \frac{1}{3} g^2 \sum_{n=1}^{\infty } \frac{c_{n}' (g)}{\epsilon^n}  \right)
\label{poles}
\end{equation}
We can now set $Z_\psi = 1$ in this last equation, working to leading order in the $1/N$ expansion. The term free of poles in the $\epsilon$ variable leads to the result
\begin{equation}
\gamma_d = - \frac{1}{N} g c_1' (g)
\end{equation}
However, this identification makes sense only if all the poles cancel out at the right-hand-side of Eq. (\ref{poles}), which implies the consistency conditions
\begin{equation}
c_{n+1}' (g)  - \frac{1}{3} g c_n' (g) = 0
\label{hi}
\end{equation}
Quite remarkably, it can be seen that the analytic evaluation of the residues $c_n (g)$ provides expressions like those in (\ref{cs0})-(\ref{cs}) that satisfy the hierarchy (\ref{hi}). We have also checked that the numerical computation of the residues, carried out to much higher order in perturbation theory, leads to power series representations of $c_n (g)$ for which (\ref{hi}) holds (at least up to order $g^{32}$). 

\begin{figure}[h]
\begin{center}
\includegraphics[width=7cm]{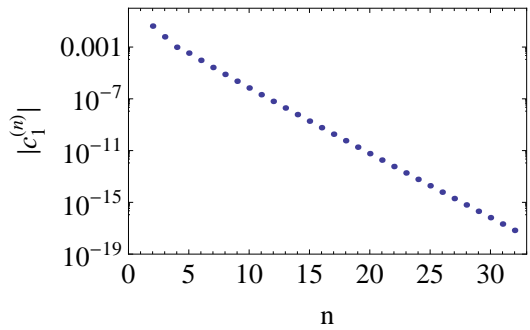}
\end{center}
\caption{Plot of the absolute value of the coefficients $c_1^{(n)}$ in the expansion of $c_1(g)$ as a power series of the coupling $g$.}
\label{four}
\end{figure}

The reiterated use of (\ref{reslt}) affords indeed a deeper numerical investigation of the residues $c_n (g)$. Thus, we have been able to obtain the coefficients of the expansion
\begin{equation}
c_1 (g )  = \sum_{n=1}^{\infty}  c_1^{(n)} g^n
\label{geom}
\end{equation}
up to the mentioned order $g^{32}$. As evidenced from the results represented in Fig. \ref{four}, the expansion (\ref{geom}) approaches a geometric series at large $n$, which means that it must have a finite radius of convergence in the variable $g$. An excellent fit of the $n$-dependence of $c_1^{(n+1)}/c_1^{(n)}$ is achieved by assuming the scaling behavior
\begin{equation}
\frac{c_1^{(n)}}{c_1^{(n-1)}} = r + \frac{r'}{n} + \frac{r''}{n^2} + \frac{r'''}{n^3} + \ldots
\end{equation} 
We get in this way an accurate estimate
\begin{equation}
r \approx 0.3333333
\label{crit}
\end{equation}
leading to the radius of convergence $g_c = 1/r \approx 3.0 \pm 1.0 \times 10^{-7}$.

A similar approach can be used to obtain the expansion of the residues $b_n(g)$ in the renormalization factor $Z_v$ for the Fermi velocity. In this case, the analytic computation of the first perturbative orders leads to the expressions
\begin{eqnarray}
b_1 (g ) & = & - c_1 (g ) -  \frac{1}{3} g
    -  \frac{1}{72} g^2  -  \frac{1}{324}  g^3 
    -  \left( \frac{1}{1728}+\frac{\zeta (3)}{1296}  \right)  g^4 
   -   \left(  \frac{1}{9720}+\frac{\pi ^4}{583200}+\frac{\zeta (3)}{19440} \right)
                                  g^5          +  \ldots     \label{first}  \\
b_2 (g ) & = &  - c_2 (g )  - \frac{1}{18} g^2  - \frac{1}{324} g^3 -  \frac{1}{1296}  g^4  
  -  \left(  \frac{1}{6480}+\frac{\zeta (3)}{4860}   \right)   g^5   +  \ldots   \\
b_3 (g ) & = &  - c_3 (g )  - \frac{1}{81}  g^3  -  \frac{1}{1296} g^4 -  \frac{1}{4860} g^5  +  \ldots      \\
b_4 (g ) & = &  - c_4 (g ) - \frac{1}{324} g^4  -  \frac{1}{4860}   g^5    +  \ldots                      \\
b_5 (g ) & = &  - c_5 (g ) - \frac{1}{1215}  g^5  +   \ldots  
\label{last}
\end{eqnarray}

The residue $b_1 (g)$ is particularly important, since it gives the rate of variation of the Fermi velocity with respect to changes in the energy scale. This can be shown by noticing that the bare Fermi velocity $Z_v v_F$ cannot depend on $\mu $, since the unrenormalized theory does not know about that auxiliary scale. We have therefore
\begin{equation}
\frac{\mu }{v_F} \frac{\partial }{\partial \mu }  \left(  Z_v v_F  \right)  =  0
\label{sfree}
\end{equation}
The dependence on $\mu $ comes only from its appearance in the definition of the renormalized coupling $g$, so that we get
\begin{equation}
Z_v \frac{\mu }{v_F} \frac{\partial }{\partial \mu } v_F + \mu \frac{\partial g}{\partial \mu } \frac{1}{N} \sum_{n=1}^{\infty } \frac{b_{n}' (g)}{\epsilon^n} = 0
\label{pws}
\end{equation}
We can now use Eq. (\ref{scalg}) to write (\ref{pws}) in the form
\begin{equation}
Z_v \frac{\mu }{v_F} \frac{\partial }{\partial \mu } v_F - \frac{1}{N} g \sum_{n=0}^{\infty } \frac{b_{n+1}' (g)}{\epsilon^n} + \frac{1}{N} \frac{1}{3} g^2 \sum_{n=1}^{\infty } \frac{b_{n}' (g)}{\epsilon^n}  = 0
\label{poles2}
\end{equation}
Assuming that the renormalized Fermi velocity $v_F$ must be free of poles in the $\epsilon $ variable, we get to leading order in the $1/N$ expansion
\begin{equation}
\frac{\mu }{v_F} \frac{\partial }{\partial \mu } v_F = \frac{1}{N} g b_1' (g)
\end{equation}
The finiteness of $\partial v_F /\partial \mu $ requires indeed the cancellation of all the poles in Eq. (\ref{poles2}), which is enforced by the set of conditions
\begin{equation}
b_{n+1}' (g)  - \frac{1}{3} g b_n' (g) = 0
\label{hip}
\end{equation}
As in the case of the anomalous scaling dimension, it can be checked that the constraints (\ref{hip}) are now satisfied by the power series expansions of the residues $b_n (g)$ obtained from our perturbative computation.

From a practical point of view, it can be seen that the coefficients in expansions like those in (\ref{first})-(\ref{last}) do not have a constant sign within each of the power series. A more extensive numerical computation, carried out to high orders in the $g$ coupling, shows actually that the residues $b_n (g)$ are regular functions in the range up to the critical $g_c$ obtained above. This implies in particular that the scaling of the Fermi velocity has a smooth behavior in the 3D Dirac semimetals, as illustrated with the plot of $g b_1' (g)$ (that we have computed to order $g^{28}$) in Fig. 1 of the main text.

\end{document}